\documentclass[conference]{IEEEtran}
\IEEEoverridecommandlockouts
\usepackage{cite}
\usepackage{amsmath,amssymb,amsfonts}
\usepackage{algorithmic}
\usepackage{graphicx}
\usepackage{textcomp}
\usepackage{xcolor}

\usepackage{bm}
\usepackage{subfigure}
\usepackage{algorithmic}
\usepackage[lined,boxed,commentsnumbered, ruled]{algorithm2e}
\newcommand{\mini}{\sf{minimize}}

\def\BibTeX{{\rm B\kern-.05em{\sc i\kern-.025em b}\kern-.08em
    T\kern-.1667em\lower.7ex\hbox{E}\kern-.125emX}}
\begin{document}

\title{Byzantine-Resilient Federated Machine Learning via Over-the-Air Computation\thanks{This work was supported in part by the National Natural Science Foundation of China (NSFC) under Grant $62001294$.}}

\author{\IEEEauthorblockN{$\text{Shaoming Huang}^{*\dag\ddagger}$, $\text{Yong Zhou}^{*}$, $\text{Ting Wang}^{\S}$, and $\text{Yuanming Shi}^{*}$}
\IEEEauthorblockA{${}^{*}$\text{School of Information Science and Technology, ShanghaiTech University, Shanghai $201210$, China} \\
${}^{\dag}$\text{Shanghai Institute of Microsystem and Information Technology, Chinese Academy of Sciences, China}\\
${}^{\ddagger}$\text{University of Chinese Academy of Sciences, Beijing $100049$, China}\\
${}^{\S}$\text{Shanghai Key Lab. of Trustworthy Computing, School of Software Engineering, East China Normal University}\\
E-mail: \{huangshm, zhouyong, shiym\}@shanghaitech.edu.cn, twang@sei.ecnu.edu.cn}
}

\maketitle

\begin{abstract}
Federated learning (FL) is recognized as a key enabling technology to provide intelligent services for future wireless networks and industrial systems with delay and privacy guarantees. However, the performance of wireless FL can be significantly degraded by Byzantine attack, such as data poisoning attack, model poisoning attack and free-riding attack. To design the Byzantine-resilient FL paradigm in wireless networks with limited radio resources, we propose a novel communication-efficient robust model aggregation scheme via over-the-air computation (AirComp). This is achieved by applying the Weiszfeld algorithm to obtain the smoothed geometric median aggregation against Byzantine attack. The additive structure of the Weiszfeld algorithm is further leveraged to match the signal superposition property of multiple-access channels via AirComp, thereby expediting the communication-efficient secure aggregation process of FL. Numerical results demonstrate the robustness against Byzantine devices and good learning performance of the proposed approach.
\end{abstract}

\section{Introduction}
With the explosive growth of modern science and technology, massive data which are critical for machine learning tasks can be exploited to improve user experience \cite{letaief2019roadmap}, such as next-word-prediction \cite{hard2018federated} and image recognition. Nevertheless, the locally generated data are often enormous and privacy-sensitive, which may preclude model training with the conventional cloud-centric approach \cite{niknam2019federated, shi2020communication, yang2020communication}. Instead, an emerging decentralized approach called Federated learning (FL) \cite{mcmahan2017communication, kairouz2019advances} has attracted considerable attention in recent years. FL avoids the direct raw data exchange among edge devices to alleviate the privacy concerns while collaboratively training a common model under the orchestration of a central server. However, a number of challenges arise for the practical deployment of FL, including the statistical challenges with non-IID (not independent and identically distributed) datasets across edge devices \cite{yang2020federated, mcmahan2017communication, lim2020federated}, high communication costs during the training process \cite{yang2020federated, mcmahan2017communication}, privacy and security concerns because of adversarial devices \cite{lim2020federated, dong2019secure, niknam2019federated}, heterogeneous devices with varying resource constraints \cite{lim2020federated}, and system design issues \cite{bonawitz2019towards}, such as the unreliable device connectivity, interrupted execution and slow convergence.

Although the raw data are not required to be sent to the cloud server, privacy and security concerns may still arise when the devices and/or the server are curious and malicious. As specified in \cite{ateniese2015hacking, tramer2016stealing}, transmitting local model parameters during the training process can still reveal sensitive information to an adversary. In order to preserve privacy, the authors in \cite{abadi2016deep} proposed a differentially privacy method, where the key idea is to introduce random perturbation to the local models before transmission. Furthermore, the authors in \cite{liu2020privacy} showed that, when FL is implemented in wireless systems via uncoded transmission, the channel noise can directly act as a privacy-inducing mechanism. In practice, there may be security issues from corrupted devices in the FL system, which send arbitrary messages due to hardware/software malfunction or malicious adversary. The adversarial devices are referred to as Byzantine devices \cite{pmlr-v80-yin18a}. In general, the Byzantine attack is classified into three categories according to attack mode in \cite{lim2020federated}, i.e., data poisoning attack \cite{wang2020attack}, model poisoning attack \cite{bhagoji2019analyzing, bagdasaryan2020backdoor} and free-riding attack \cite{kim2019blockchained}.

The state-of-the-art research progress has been made on secure FL in presence of Byzantine devices. The authors in \cite{dong2019secure} classified the current secure FL algorithms into four categories: robust aggregation rule \cite{damaskinos2019aggregathor, dong2019secure}, preprocess method from the information-the-oretical perspective \cite{chen2018draco, mhamdi2018hidden}, model with regularization term \cite{li2019rsa}, and adversarial detection \cite{alistarh2018byzantine, so2020byzantine}. In this paper, we focus on the robust aggregation rule. Notably, it has been shown that even a single Byzantine fault can significantly alter the trained model with naive mean-value aggregation rule \cite{blanchard2017machine}. Therefore, multiple robust aggregating rules are explored against undesired Byzantine failures \cite{damaskinos2019aggregathor, dong2019secure}, such as geometric median \cite{minsker2015geometric, chen2017distributed, wu2020federated}, coordinate-wise median \cite{pmlr-v80-yin18a}, trimmed mean \cite{pmlr-v80-yin18a} and Krum \cite{blanchard2017machine}. In \cite{chen2017distributed} and \cite{wu2020federated}, it was shown that, when the number of Byzantine devices is less than the number of reliable devices, the distributed Stochastic Gradient Descent (SGD) with geometric median aggregation rule linearly converges to a neighborhood of the optimal solution.

\begin{figure*}[ht]
\centering{
    \subfigure[Global Model Dissemination]{\includegraphics[width=0.325\textwidth]{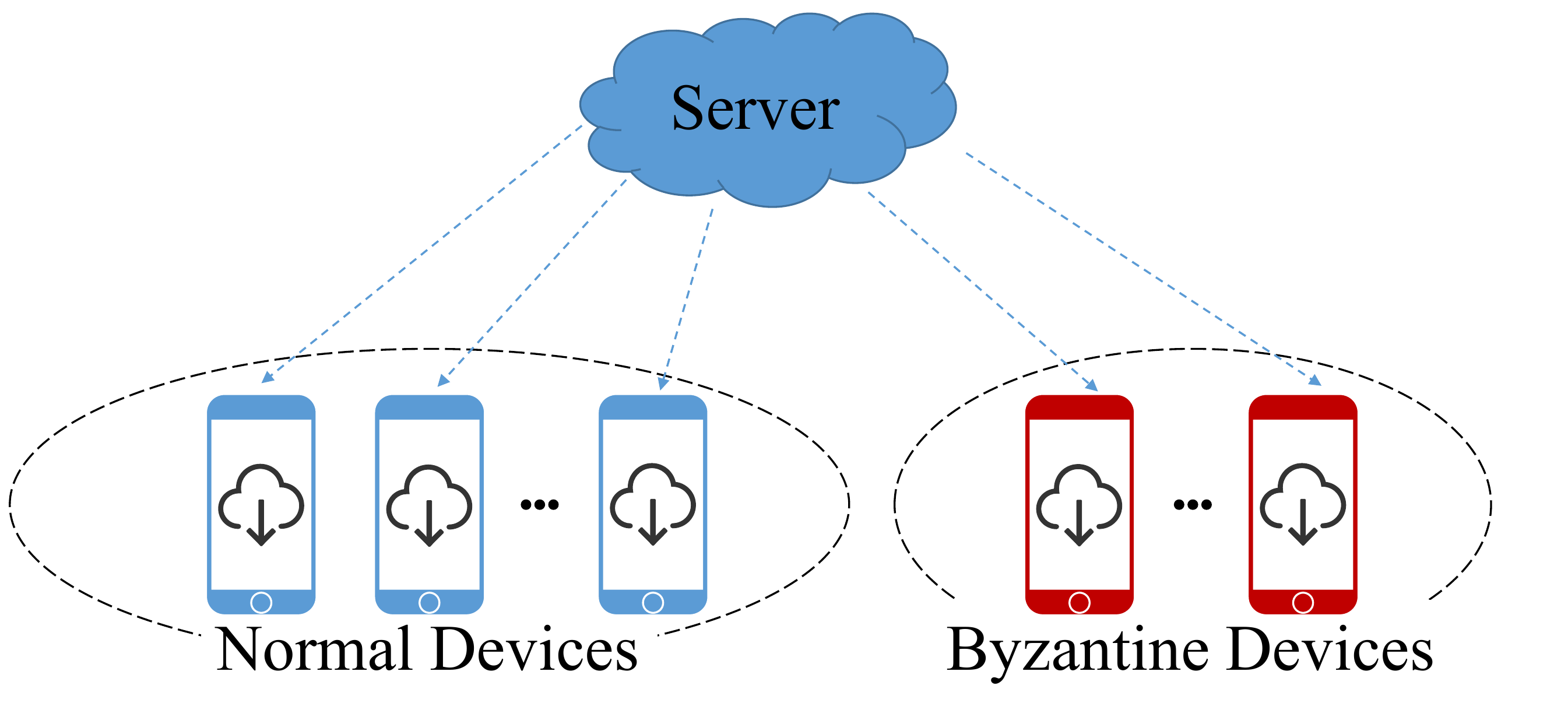}}
    \subfigure[Local Model Computation]{\includegraphics[width=0.325\textwidth]{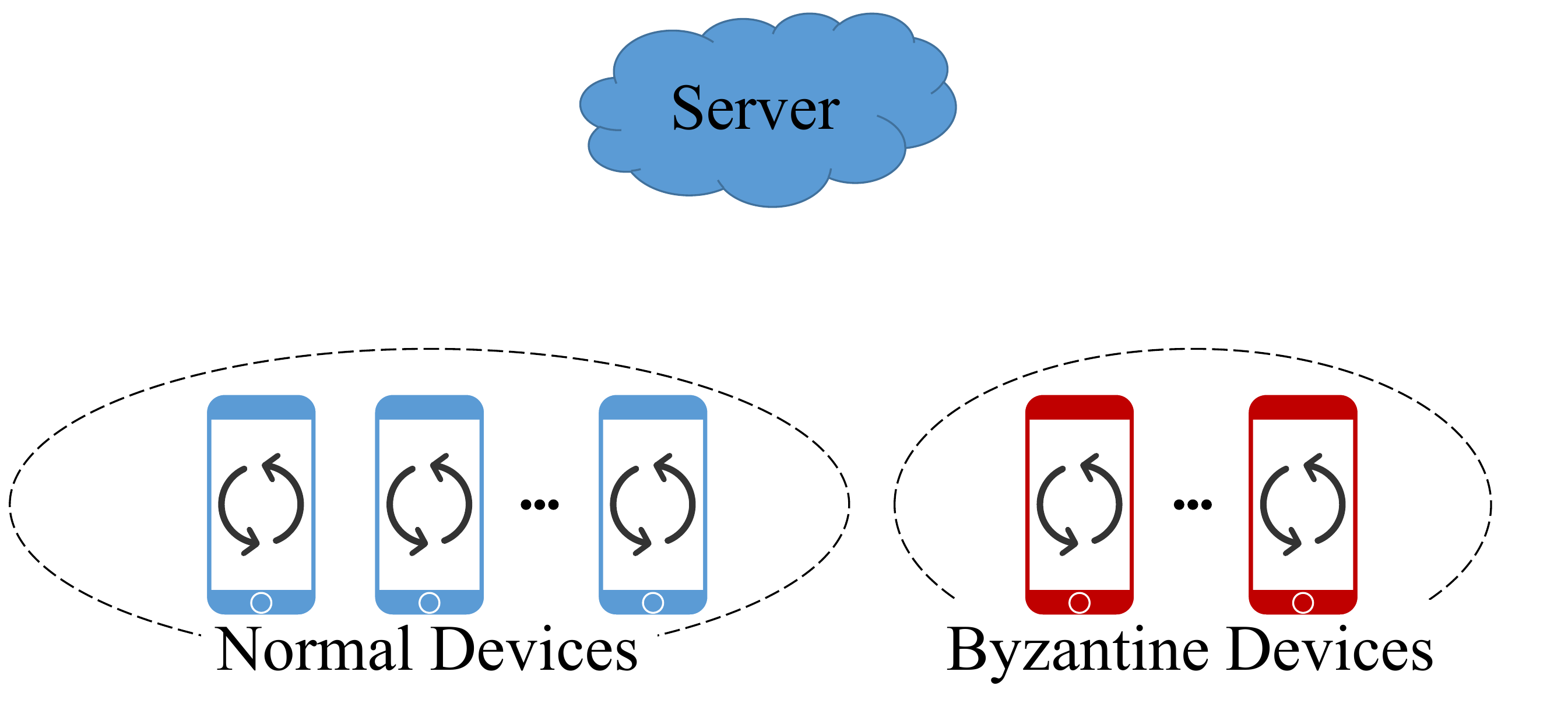}}
    \subfigure[Model Aggregation \& Update]{\includegraphics[width=0.325\textwidth]{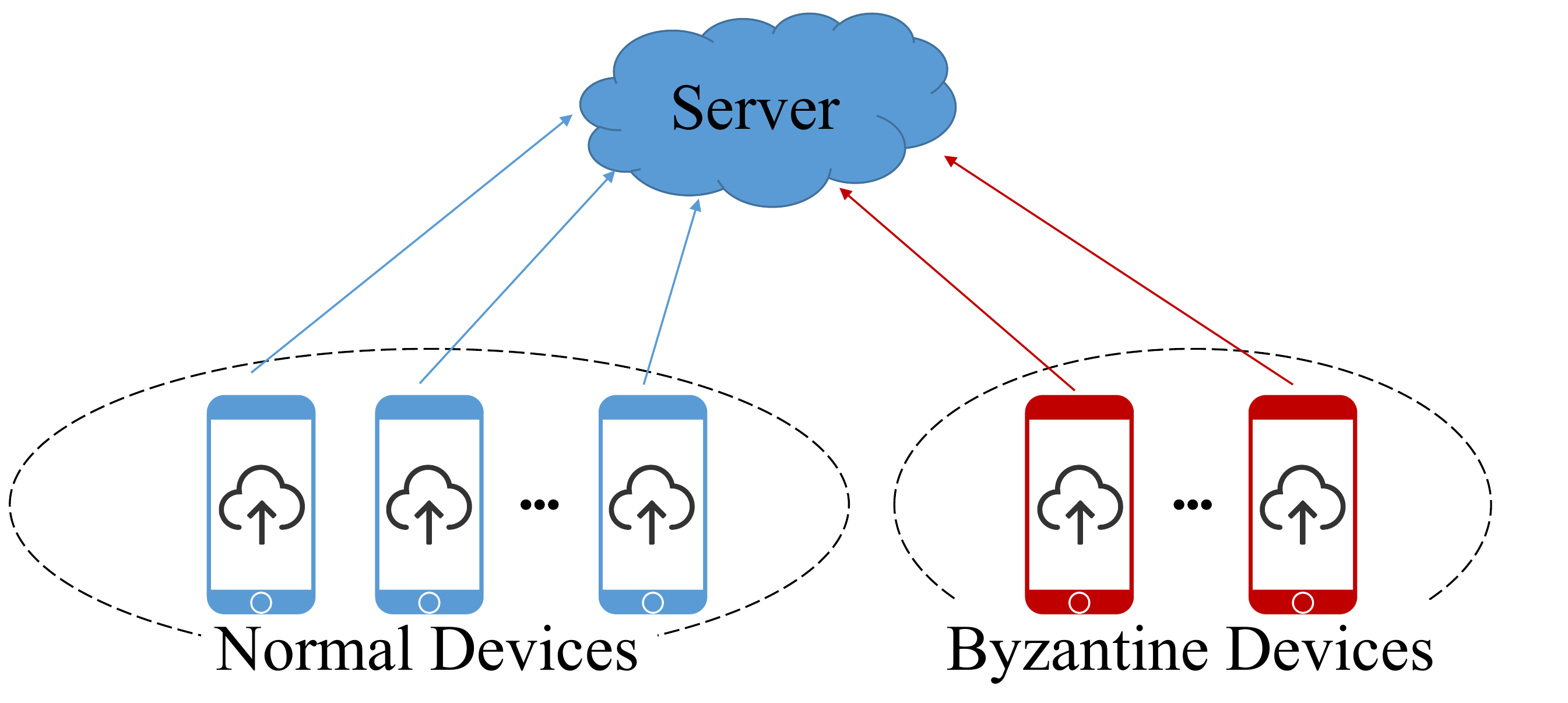}}
    \caption{One round of the distributed on-device FL system consisting of one server, normal devices and Byzantine devices.}
    \label{FL_Byzantine}
}
\end{figure*}

In spite of the good convergence guarantee, the geometric median aggregation rule requires solving a convex optimization problem to obtain the aggregation result, whose computational complexity is prohibitive when the dimension of model parameters is large. To reduce the computation complexity, we adopt the improved Weiszfeld algorithm \cite{pillutla2019robust}, which introduces a smoothed factor for numerical stability. Furthermore, since information exchange is limited by radio resources in wireless networks, we propose a novel communication-efficient robust aggregation scheme based on over-the-air computation (AirComp) \cite{nazer2007computation} to expedite the aggregation process under the wireless communication, which utilizes the signal superposition property of multiple-access channels to improve the communication efficiency and reduce the required bandwidth \cite{nazer2007computation, Yang2020FLRIS, wang2020federated, goldenbaum2013harnessing, Wang2021AirCompRIS, Dong2020BlindAirComp}. Numerical results demonstrate that the proposed approach enjoys the robustness against Byzantine attack and good learning performance under wireless environment.


\section{System Model}
\label{System_Model}
In this section, we first present the distributed on-device FL system to collaboratively learn a common model under the orchestration of a central server. Considering the existence of Byzantine devices, the FL system with a robust aggregation rule is extended to suppress the undesired bias of the learning process.

\subsection{Federated Learning Model}
We consider the distributed on-device FL system with one server and $K$ devices \cite{mcmahan2017communication}, which aims to collaboratively learn a common model (e.g., logistic regression and deep neural network). We denote the set of all devices as $\mathcal{K}$, and each device $k \in \mathcal{K}$ has its own local dataset, of which the raw data are unavailable to other devices and the server. By contrast, the server has no access to any dataset, and orchestrates the learning process via exchanging model parameters with each device. In general, we are interested in the finite-sum optimization problem \cite{pillutla2019robust}
\begin{equation}
    \underset{\bm{w}\in \mathbb{R}^{d}}{\mini} \quad F(\bm{w}) := \sum_{k \in \mathcal{K}} \alpha_{k} \mathbb{E}_{\bm{\xi} \sim \mathcal{D}_{k}} \left[ f \left( \bm{w}; \bm{\xi} \right) \right],
    \label{main_problem}
\end{equation}
where $\bm{w} \in \mathbb{R}^{d}$ represents the parameters of the common model, $\bm{\xi} \in \Xi$ represents random sample variable according to a certain probability distribution, $f : \mathbb{R}^{d} \times \Xi$ represents the loss function, for each device $k \in \mathcal{K}$, the $\mathcal{D}_{k}$ is a probability distribution supported on $\Xi$ and the weight factor $\alpha_{k} > 0$ with $\sum_{k \in \mathcal{K}} \alpha_{k} = 1$. Remarkably, the term $\alpha_{k}$ specifies the relative weight of each device, with natural setting being $\alpha_{k} = n_{k}/n$ \cite{FedSum2020}, where $n_{k}$ is the number of data samples in device $k$ and $n = \sum_{k \in \mathcal{K}} n_{k}$ is the total number of data samples. We assume that $f \left( \cdot; \bm{\xi} \right)$ is continuously differentiable for any random sample variable $\bm{\xi}$, and the expectation in \eqref{main_problem} is well-defined and finite. The goal of FL process is to obtain a common model with good performance via solving the optimization problem \eqref{main_problem} collaboratively.

In particular, in supervised machine learning, the random variable $\bm{\xi} = (\bm{x}, y)$ is a data-label pair and loss function $f(\bm{w}; \bm{\xi}) = \ell(y, \phi(\bm{x}; \bm{w}))$, where function $\phi$ maps the data $\bm{x}$ to a prediction made with model parameters $\bm{w}$, and $\ell$ is a certain loss function such as the squared loss and cross entropy. For instance, the mapping function of linear regression can be expressed as $\phi(\bm{x}, \bm{w}) = \bm{w}^{\sf{T}} \bm{x}$.

We now elucidate the canonical FL \cite{mcmahan2017communication}, which runs in synchronized rounds of computation and communication process between the server and devices, consisting of three steps:
\begin{enumerate}
\item \textit{Global Model Dissemination:} The server broadcasts the global model parameters to all devices.
\item \textit{Local Model Computation:} Each device receives the current global model parameters, executes some local computation to generate update message based on its own local dataset.
\item \textit{Model Aggregation \& Update:} All devices transmit their update messages. The server aggregates these update messages to obtain a new global model parameters according to a certain aggregation rule.
\end{enumerate}
These steps are repeated until the whole system attains a sufficiently trained model, as shown in Fig. \ref{FL_Byzantine}. Note that there are two prevalent paradigms about the update messages, i.e., stochastic gradients \cite{blanchard2017machine, chen2018draco, chen2017distributed, pmlr-v80-yin18a, damaskinos2019aggregathor, wu2020federated} and model weight parameters \cite{pmlr-v80-yin18a, pillutla2019robust, mcmahan2017communication}. In this paper, we choose the model weight parameters instead of stochastic gradients due to more efficient local computation and more reasonable distribution assumption mentioned in \cite{pillutla2019robust}.

\subsection{Byzantine Attack and Robust Aggregation}
\label{Byzantine_Attack_Aggregation}
We suppose that there are $B$ Byzantine devices among $K$ devices in the system. The Byzantine devices \cite{pmlr-v80-yin18a}, unbeknownst to the normal devices and the server, are capable of misleading the optimization process by colluding and sending elaborately malicious update messages to the server. For brevity, we define the set of all Byzantine devices as $\mathcal{B}$.

To enhance the system robustness against the Byzantine attack, multiple robust aggregation rules have been studied so far \cite{damaskinos2019aggregathor, dong2019secure}, such as geometric median \cite{minsker2015geometric, chen2017distributed, wu2020federated}, coordinate-wise median \cite{pmlr-v80-yin18a}, trimmed mean \cite{pmlr-v80-yin18a} and Krum \cite{blanchard2017machine}, which are Byzantine-resilient comparing to naive mean \cite{mcmahan2017communication}. It was shown that when the number of Byzantine devices satisfies $B < K/2$, distributed SGD with geometric median aggregation rule linearly converges to a neighborhood of the optimal solution \cite{chen2017distributed, wu2020federated}. Owing to this good convergence guarantee, we choose geometric median as the robust aggregation rule, which is to obtain a vector $\bm{z}^{\star}$ satisfying
\begin{equation}
	\bm{z}^{\star} = \operatorname{arg} \underset{ \bm{z} }{\operatorname{min}} \sum_{k \in \mathcal{K}} \alpha_{k} \|\bm{z} - \bm{w}_{k}\|,
\label{gm}
\end{equation} 
where $\alpha_{k} > 0$ is the weight factor, $\|\cdot\|$ denotes Euclidean norm function, and the objective function is denoted as $g(\bm{z}) := \sum_{k \in \mathcal{K}} \alpha_{k} \| \bm{z} - \bm{w}_{k}\|$. In essence, we expect a vector in the whole normal space $\mathbb{R}^{d}$ with minimal distance to all vectors in $\{\bm{w}_{k}, k \in \mathcal{K} \}$. We apply the Weiszfeld algorithm \cite{weiszfeld1937point} to solve the original problem \eqref{gm} iteratively,
\begin{equation}
    \bm{z}^{(t+1)} = \frac{\sum_{k \in \mathcal{K}} \beta_{k}^{(t)} \bm{w}_{k} }{\sum_{k \in \mathcal{K}} \beta_{k}^{(t)} }
\end{equation}
with 
\begin{equation}
\beta_{k}^{(t)} = \frac{\alpha_{k}}{ \|\bm{z}^{(t)} - \bm{w}_{k} \| },
\nonumber
\end{equation}
which however is numerically instable due to division by possibly extremely close distance $\|\bm{z}^{(t)} - \bm{w}_{k} \|$.

\subsection{Smoothed Geometric Median}
As a consequence, we introduce smooth factor $\nu > 0$ to avoid the extremely small denominator \cite{pillutla2019robust}. As a result, the smoothed function can be expressed as
\begin{equation}
    g_{\nu}(\bm{z}) := \sum_{k \in \mathcal{K}} \alpha_{k} \| \bm{z} - \bm{w}_{k}\|_{(\nu)},\\ 
\end{equation}
where
\begin{equation}
    \| \bm{z}\|_{(\nu)} = \left\{
    \begin{aligned}
        &\frac{1}{2 \nu} \|\bm{z}\|^{2} + \frac{\nu}{2}, &\|\bm{z}\| \leq \nu\\
        &\|\bm{z}\|, &\|\bm{z}\| > \nu
    \end{aligned}\right..
\nonumber
\end{equation}
Note that the objective function $g_{\nu}$ is a $1/\nu$-smooth approximation to $g$ \cite{Beck2012SmoothingAF}. Obtaining the smoothed geometric median $\operatorname{arg} {\min} \ g_{\nu}(\bm{z})$ is a convex optimization problem, which can also be efficiently solved via the Weiszfeld algorithm \cite{weiszfeld1937point}, i.e.,
\begin{equation}
    \bm{z}^{(t+1)} = \frac{\sum_{k \in \mathcal{K}} \beta_{k}^{(t)} \bm{w}_{k} }{\sum_{k \in \mathcal{K}} \beta_{k}^{(t)} }
\label{smoothed_GM_Weiszfeld}
\end{equation}
with
\begin{equation}
\beta_{k}^{(t)} = \frac{\alpha_{k}}{ \max\{\nu, \|\bm{z}^{(t)} - \bm{w}_{k} \|\} },
\nonumber
\end{equation}
which enjoys sublinear convergence rate \cite{pillutla2019robust}. In practice, we choose the $\bm{w}^{(t)}$ as the initial iteration point of Weiszfeld algorithm, and the Byzantine-resilient FL with the smoothed geometric median aggregation rule is listed in \textbf{Algorithm \ref{FL_GM}}. Note that the aggregation step of federated average approach in \cite{mcmahan2017communication} has just one average operation, which is vulnerable to Byzantine attack due to the naive mean-value aggregation rule. By contrast, in \textbf{Algorithm \ref{FL_GM}}, we need several communication rounds to find the smoothed geometric median for defending against Byzantine attack, which requires high computation and communication costs. To tackle this challenge, we propose a communication-efficient secure aggregation approach based on AirComp. Since each iteration of the Weiszfeld algorithm can be seen as taking a weighted average of the $\bm{w}_{k}$'s, we can harness the signal superposition property of multiple-access channels to enable fast robust aggregation in FL, which will be discussed in the next section.

\begin{algorithm}
\label{FL_GM}
\caption{Ideal Byzantine-Resilient FL with Smoothed Geometric Median Aggregation Rule}
\KwIn{$F$ from \eqref{main_problem}, step size $\eta$}
\For{$t = 0, 1, \cdots$}
{
    Broadcast $\bm{w}^{(t)}$ to each device $k \in \mathcal{K}$.\\
    \For{\rm each device $k \in \mathcal{K}$ \textbf{in parallel}}
    {
        $\bm{w}_{k}^{(t)} \leftarrow $ \textit{LocalComp}$(k, \bm{w}^{(t)})$
    }
    \textbf{Initialization:} $\bm{z} = \bm{w}^{(t)}$ \qquad $\triangleleft$ Aggregation\\
    \Repeat{\rm $\bm{z}$ \textbf{converges}}
    {
        Broadcast $\bm{z}$ to each device $k \in \mathcal{K}$.\\
        \For{\rm each device $k \in \mathcal{K}$ \textbf{in parallel}}
        {
            $\beta_{k} = {\alpha_{k}}/{ \max\{\nu, \|\bm{z} - \bm{w}_{k}^{(t)} \|\} }$
        }
        $\bm{z} = \frac{\sum_{k \in \mathcal{K}} \beta_{k} \bm{w}_{k}^{(t)} }{\sum_{k \in \mathcal{K}} \beta_{k} }$
    }
    $\bm{w}^{(t+1)} \leftarrow \bm{z}$ \qquad $\triangleleft$ Global model update\\
}
\end{algorithm}

\section{communication-Efficient Model Aggregation Protocol}
\label{Communication_Protocol}
In this section, we present the implement of Byzantine-resilient FL in wireless networks and investigate the impact of wireless channel and noise on the resilience and learning performance. Utilizing the additive structure of model aggregation in \eqref{smoothed_GM_Weiszfeld}, we propose a computation and communication co-designed approach to enhance the overall system efficiency based on the principle of AirComp.

We focus on the information exchange process between the server and devices in shared wireless channels. For simplicity, we consider the multi-user single-input single-output (SISO) wireless communication system, where all the server and devices are equipped with single antenna. To account for the fact that the base station has less stringent power constraint than devices, we assume that the downlink communication from the server to devices is ideal \cite{liu2020privacy} and each device can receive the current global model parameters without any distortion. Instead, we pay attention to the uplink transmission from devices to the server. To further simplify the communication model, we assume that the channel coefficient remains constant with a communication block and varies over successive blocks with potential correlation \cite{liu2020privacy}. During a communication block, we expect that each device transmits entire update message with dimension $m$ to the server. This assumption is practically well justified when on-device FL models are typically light-weight under a few tens of thousands of parameters \cite{ravi2019efficient}, whose model weight parameters dimension is in the same order of coherence block's magnitude \cite{wang2018doppler}. Therefore, it is reasonable to communicate the entire update message within one communication block for each device \cite{liu2020privacy}.

We propose an AirComp based scheme to support update message transmission, where all devices concurrently transmit uncoded local model parameters to the server. The received signal at the server during the $t$-th block is expressed as
\begin{equation}
    \boldsymbol{y}^{(t)} = \sum_{k \in \mathcal{K} }h_{k}^{(t)} \boldsymbol{x}_{k}^{(t)} + \boldsymbol{n}^{(t)},
\label{NOMA_received}
\end{equation}
where for device $k$ at iteration $t$, $h_{k}^{(t)} \in \mathbb{C}$ is the channel coefficient, $\boldsymbol{x}_{k}^{(t)} \in \mathbb{C}^{m}$ is an uncoded function about the local update message, and channel noise $\boldsymbol{n}^{(t)} \in \mathbb{C}^{m}$ is an i.i.d. random variable according to $\mathcal{CN}(\boldsymbol{0}, \sigma^{2}\boldsymbol{I})$. Because the transmit power of device is usually limited, we constrain the uncoded function $\boldsymbol{x}_{k}^{(t)}$ with
\begin{equation}
    \mathbb{E} \left[ \| \boldsymbol{x}_{k}^{(t)} \|^{2} \right] \leq m \times P.
\label{power_constrain}
\end{equation}

As in most existing works on AirComp \cite{sery2020analog, amiri2020machine, yang2020federated, liu2020privacy}, we assume perfect channel state information (CSI) is available at all devices, so that the phase shift of each channel can be compensated at the transmitter, ensuring the real and non-negative channel gain for each device $k$ at iteration $t$. To this end, we apply the channel inversion at device $k$ \cite{liu2020privacy}
\begin{equation}
\boldsymbol{x}_{k}^{(t)} = \rho_{k}^{(t)} \boldsymbol{x}_{k}^{\prime (t)} \text{ with } \boldsymbol{x}_{k}^{\prime (t)} = \frac{ {h_{k}^{(t)}}^{*} \boldsymbol{m}_{k}^{(t)} }{ \|h_{k}^{(t)}\|^2 },
\label{channel_inversion}
\end{equation}
where $\boldsymbol{m}_{k}^{(t)} \in \mathbb{R}^{m}$ is update message about variables $\beta_{k}$ and $\bm{w}_{k}$ in \eqref{smoothed_GM_Weiszfeld}, ${}^*$ denotes complex conjugate operation, $\boldsymbol{x}_{k}^{\prime (t)}$ is channel inversed message without power scaling, and $\rho_{k}^{(t)} \in \mathbb{R}$ is a power scaling factor guaranteeing power constraint \eqref{power_constrain}. Substituting \eqref{channel_inversion} into \eqref{NOMA_received}, the received signal can be simplified as
\begin{equation}
\boldsymbol{y}^{(t)} = \sum_{k \in \mathcal{K} } h_{k}^{(t)} \rho_{k}^{(t)} \frac{ {h_{k}^{(t)}}^{*} \boldsymbol{m}_{k}^{(t)} }{ \|h_{k}^{(t)}\|^2 } + \boldsymbol{n}^{(t)} = \sum_{k \in \mathcal{K} } \rho_{k}^{(t)} \boldsymbol{m}_{k}^{(t)} + \boldsymbol{n}^{(t)}.
\label{NOMA_channel}
\end{equation}

We aim to expedite the aggregation process of the smoothed geometric median under wireless environment. The basic idea is utilizing the additive structure in \eqref{smoothed_GM_Weiszfeld} to realize the AirComp. In the following, we explain some implementation improvements in practice.
\begin{enumerate}
    \item We directly transmit the update message $[\beta_{k} \bm{w}_{k}^{(t)}, \beta_{k}] \in \mathbb{R}^{m}$ with $m=d+1$, because the additive structure is spread over numerator and denominator of \eqref{smoothed_GM_Weiszfeld}. However, the average power expectations $\mathbb{E} [ \|\beta_{k} \bm{w}_{k}^{(t)} \|^{2}] / d$ and $\mathbb{E} [ | \beta_{k} |^{2} ]$ may be quite different, leading to extremely uneven distribution of energy. An alternative strategy is to transmit $[\beta_{k} \bm{w}_{k}^{(t)}, \beta_{k} \sqrt{\| \bm{w}_{k}^{(t)} \|^2 / d}]$ via introduce a scaling factor $\sqrt{\| \bm{w}_{k}^{(t)} \|^2 / d}$. However, AirComp requires the received signal from different devices to achieve \textit{magnitude alignment} \cite{zhu2020over}, i.e., all scaling factors should be consistent. As a remedy, we transmit $[\beta_{k} \bm{w}_{k}^{(t)}, \beta_{k} \sqrt{ \|\bm{z}\|^2 / d}]$, since $\bm{z}$ is the information broadcasted by the server at the begin of each iteration. The change of denominator is eliminated at the server in the end of each iteration.


    \item During AirComp, the received signal from different devices should achieve \textit{magnitude alignment} at the receiver \cite{zhu2020over}, i.e., the power scaling factor $\rho_{k}^{(t)}$ should be equal for all devices in principle. Through observation, once $\|\bm{z} - \bm{w}_{k}^{(t)}\| < \nu$ and/or wireless channel $h_{k}$ is in deep fade, $\bm{x}_{k}^{\prime}$ can be extremely large because of the tiny denominator $\nu$ and/or $\|h_{k}\|^{2}$, which results in other devices' weak links to the server with low signal to noise ratio (SNR). To address this problem, we set a hard threshold $C$ to truncate the devices whose magnitude exceeds the threshold. Whereas, it will decrease the degree of devices participation, so we employ total access strategy with soft threshold $\max\{C, \|\bm{x}_{k}^{\prime}\|^{2}/m\}$ instead of the truncation strategy with hard threshold $C$. For device $k$ with $\|\bm{x}_{k}^{\prime}\|^{2}/m > C$, the transmitted message will become distorted, which is termed transmission distortion. Hence, the threshold $C$ should be properly adjusted to balance a trade-off between the transmission distortion and SNR.
    \item At the server, we isolate the real part from received signal $\bm{y}$, since the phase shift of each channel has been compensated. And we multiply $\sqrt{\|\bm{z}\|^2/d}$ to eliminate denominator change from the devices.
\end{enumerate}

The Weiszfeld algorithm for smoothed geometric median aggregation rule based on AirComp are elaborated in \textbf{Algorithm \ref{NOMA_GM}}. Furthermore, the whole process of Byzantine-resilient FL paradigm via AirComp is summarized in \textbf{Algorithm \ref{FL_AIR}}.

\begin{algorithm}
\label{NOMA_GM}
\caption{Weiszfeld Algorithm for Smoothed Geometric Median via AirComp}
\KwIn{Initialization $\bm{z}$}
\Repeat{\rm $\bm{z}$ \textbf{converges}}
{
    Broadcast $\bm{z}$ to each device $k \in \mathcal{K}$.\\
    \For{\rm each device $k \in \mathcal{K}$ \textbf{in parallel}}
    {
        $\beta_{k} = {\alpha_{k}}/{ \max\{\nu, \|\bm{z} - \bm{w}_{k}^{(t)} \|\} }$\\
        $\bm{m}_{k} = [\beta_{k}  \bm{w}_{k}^{(t)}, \beta_{k}  \sqrt{\|\bm{z}\|^2/d}]$\\
        $\bm{x}_{k}^{\prime} = { {h_{k}}^{*} \boldsymbol{m}_{k} } / { \|h_{k}\|^2 }$\\
        \textbf{Power Control:}\\
        $\bm{x}_{k} = \rho_{k} \bm{x}_{k}^{\prime}$ with $\rho_{k} = \sqrt{ \frac{P}{\max\{C, \|\bm{x}_{k}^{\prime}\|^{2}/m\}} } $
    }
    $[\bm{a}, b] = \operatorname{Real}\{\bm{y}\}$, where $\bm{y}$ is obtained via \eqref{NOMA_received}\\
    $\bm{z} \leftarrow \bm{a} / b * \sqrt{\|\bm{z}\|^2/d}$
}
\textbf{return} $\bm{z}$
\end{algorithm}

\begin{algorithm}
\label{FL_AIR}
\caption{Byzantine-resilient FL Paradigm via AirComp}
\KwIn{$F$ from \eqref{main_problem}, step size $\eta$}
\For{$t = 0, 1, \cdots$}
{
    Broadcast $\bm{w}^{(t)}$ to each device $k \in \mathcal{K}$.\\
    \For{\rm each device $k \in \mathcal{K}$ \textbf{in parallel}}
    {
        $\bm{w}_{k}^{(t)} \leftarrow $ \textit{LocalComp}$(k, \bm{w}^{(t)})$
    }
    \textbf{Initialization:} $\bm{z} = \bm{w}^{(t)}$ \qquad $\triangleleft$ Aggregation\\
    Obtain $\bm{z}$ via \textbf{Algorithm \ref{NOMA_GM}}\\
    $\bm{w}^{(t+1)} \leftarrow \bm{z}$ \qquad $\triangleleft$ Global model update\\
}
\end{algorithm}

\begin{figure*}[ht]
    \centering
    \includegraphics[width=1.0\textwidth]{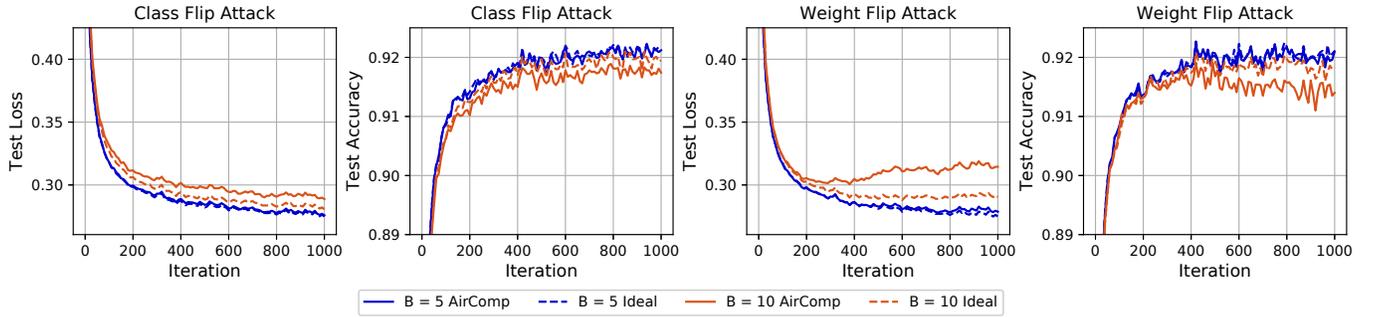}
    \caption{Comparison convergence of the AirComp-aided and the ideal Byzantine-resilient FL under different number of Byzantine devices, with channel noise variance $\sigma^{2} = 10^{-2}$ and threshold $C = 500 \times \|\bm{z}\|^{2}/m$. "AirComp" represents the wireless Byzantine-resilient FL based on AirComp, and "Ideal" represents the ideal Byzantine-resilient FL without wireless transmission. The left two plots show the convergence under class flip attack, and the right two plots show the convergence under weight flip attack.}
    \label{attack_pattern}
\end{figure*}

\section{Simulation Results}
\label{Simulation_Experiment}
In this section, we provide simulation results to evaluate the performance of the proposed AirComp assisted Byzantine-resilient FL framework.

\subsection{Simulation Setup}
Our simulation results are conducted using pytorch $1.1.0$ under python $3.7.4$ environment. We consider the handwritten-digit recognition task on MNIST dataset, whose sample has $28 \times 28$ features and belongs to $10$ different classes, i.e., label from $0$ to $9$. We partition the $60,000$ training data into $K = 50$ mutually independent subsamples with equal size at random, whose weight factor $\alpha_{k}$ is $\frac{1}{K}$. We adopt the multi-class logistic regression model and the \textit{LocalComp} function in \textbf{Algorithm \ref{FL_GM}} and \textbf{Algorithm \ref{FL_AIR}} is set as one-step batch-SGD optimization with batch $b = 50$ and learning rate $10^{-2}$. In the inner loop with the Weiszfeld algorithm, we set the maximal number of iterations to be $1000$, the interruption tolerance to be $10^{-5}$, and the smooth factor to be $\nu = 10^{-4}$. For uplink transmission, we generate the wireless channel $h_{k} \sim \mathcal{CN}(0,1)$ independently for device $k$, and wireless noise $\bm{n} \sim \mathcal{CN}(\bm{0}, \sigma^{2} \bm{I})$. We set the maximum transmit power of all devices to be $P = 1$.

In addition, we consider two typical Byzantine attack patterns in $\textit{Local Model Computation}$ step of malicious Byzantine devices.
\begin{enumerate}
    \item \textit{Class flip \cite{pmlr-v80-yin18a}:} It is a kind of data poisoning attack \cite{wang2020attack}. Concretely, each of the training label $i$ on the Byzantine devices is replaced with $9-i$. For example, the data sample of digit number $1$ are maliciously modified as number $8$.
    \item \textit{Weight flip \cite{pillutla2019robust}:} It is a kind of model poisoning attack \cite{bhagoji2019analyzing, bagdasaryan2020backdoor}. Concretely, the update message $\bm{w}_{l}^{\prime}$ transmitted by Byzantine device $l \in \mathcal{B}$ is modified so that the weighted arithmetic mean of the model parameters is set to be the negative of what it would be without the Byzantine attack. Mathematically, $\bm{w}_{l}^{\prime} = - \bm{w}_{l} - \frac{2}{K-B}\sum_{k \notin \mathcal{B}} \bm{w}_{k}$, for $l \in \mathcal{B}$.
\end{enumerate}

\subsection{Performance Evaluation}

We compare the wireless Byzantine-resilient FL based on AirComp with the ideal Byzantine-resilient FL without wireless communication in the simulation. The former is implemented by following \textbf{Algorithm \ref{FL_AIR}}, and the latter is implemented by following \textbf{Algorithm \ref{FL_GM}}. Herein, we set the channel noise variance $\sigma^{2} = 10^{-2}$ and threshold $C = 500 \times \|\bm{z}\|^{2}/m$ to balance a trade-off between the transmission distortion and SNR. Fig. \ref{attack_pattern} shows that the convergence of the AirComp-aided and the ideal Byzantine-resilient FL under different number of Byzantine devices with class flip and weight flip attack pattern. According to the numerical results, the AirComp-aided and the ideal FL with the smoothed geometric median robust aggregation rule both effectively resist the attack from Byzantine devices, and the wireless environment slightly degrade the performance of the Byzantine-resilient FL. The performance degradation is more distinct as the number of Byzantine devices increases. As a result, we can draw a conclusion that the proposed Byzantine-resilient FL paradigm via AirComp is robust against Byzantine devices, and can attain a satisfactory model with acceptable performance degradation compared with ideal Byzantine-resilient FL without regard to wireless transmission process.

\section{Conclusion}
\label{Conclusion}
In this paper, we proposed a novel Byzantine-resilient FL paradigm via AirComp in wireless communication system, which is communication-efficient. The smoothed geometric median aggregation rule is applied to resist the attack from Byzantine devices, and AirComp is adopted to speed up the aggregation process of wireless FL. More concretely, we employed Weiszfeld algorithm to obtain the smoothed geometric median, and further put forth computation and communication co-designed approach based on AirComp via utilizing additive structure of the Weiszfeld algorithm. The numerical results demonstrated that the proposed Byzantine-resilient FL paradigm via AirComp is able to attain a satisfactory performance with the impact of adversarial Byzantine attack and wireless environment.

\bibliography{Reference} 
\bibliographystyle{ieeetr}

\end{document}